%% file: main.tex
\documentclass[10pt,conference]{IEEEtran}
\IEEEoverridecommandlockouts
\usepackage{cite}
\usepackage{amsmath,amssymb,amsfonts}
\usepackage{algorithmic}
\usepackage{graphicx}
\usepackage{svg}
\usepackage{textcomp}
\usepackage{xcolor}
\def\BibTeX{{\rm B\kern-.05em{\sc i\kern-.025em b}\kern-.08em
    T\kern-.1667em\lower.7ex\hbox{E}\kern-.125emX}}
\begin{document}

\newcommand{\lc}[1]{\footnote{\textit{Leon:} #1.}}
\newcommand{\TODO}[1]{\textcolor{red} {$\blacktriangleright$ \textbf{#1}}}
\newcommand\kn[1]{\textcolor{red}{KN: #1}}

\title{Towards Code Generation from BDD Test Case Specifications: A Vision}

\author{\IEEEauthorblockN{Leon Chemnitz}
\IEEEauthorblockA{\textit{Software Technology Group} \\
\textit{Technische Universität Darmstadt}\\
Darmstadt, Germany \\
leon.chemnitz@stud.tu-darmstadt.de}
\and
\IEEEauthorblockN{David Reichenbach}
\IEEEauthorblockA{\textit{Software Technology Group} \\
\textit{Technische Universität Darmstadt}\\
Darmstadt, Germany \\
david.reichenbach@stud.tu-darmstadt.de}
\and
\IEEEauthorblockN{Hani Aldebes}
\IEEEauthorblockA{\textit{Software Technology Group} \\
\textit{Technische Universität Darmstadt}\\
Darmstadt, Germany \\
hani.aldebes@stud.tu-darmstadt.de}
\and
\IEEEauthorblockN{Mariam Naveed}
\IEEEauthorblockA{\textit{Software Technology Group} \\
\textit{Technische Universität Darmstadt}\\
Darmstadt, Germany \\
mariam.naveed@stud.tu-darmstadt.de}
\and
\IEEEauthorblockN{Krishna Narasimhan}
\IEEEauthorblockA{\textit{Software Technology Group} \\
\textit{Technische Universität Darmstadt}\\
Darmstadt, Germany \\
kri.nara@st.informatik.tu-darmstadt.de}
\and
\IEEEauthorblockN{Mira Mezini}
\IEEEauthorblockA{\textit{Software Technology Group} \\
\textit{Technische Universität Darmstadt}\\
Darmstadt, Germany \\
mezini@informatik.tu-darmstadt.de}
}

\maketitle

\begin{abstract}
\input{abstract}
\end{abstract}


\section{Introduction}
\input{introduction}

\section{Background}
\label{sec:background}
\input{background}


\section{Method}
\label{sec:idea}
\input{method}

\section{Discussion on the impact of our vision}
\label{sec:discussion}
\input{impact_analysis}

\section{Conclusion}
\label{sec:conclusion}
\input{conclusion}

\section{Acknowledgement}
\label{sec:acknowledgement}
\input{acknowledgement}

\bibliographystyle{IEEEtran}
\bibliography{references}

\end{document}

%% file: abstract.tex
Automatic code generation has recently attracted large attention and is becoming
more significant to the software development process.
%
%
Solutions based on Machine Learning and Artificial Intelligence are being used
to increase human and software efficiency in potent and innovative ways.
In this paper, we aim to leverage these developments and introduce a novel approach
to generating frontend component code for the popular Angular framework.
We propose to do this using behavior-driven development test specifications as
input to a transformer-based machine learning model; however, we do not provide any
proof-of-concept solution in this work.
Our approach aims to drastically reduce the development time needed for web applications while potentially
increasing software quality and introducing new research ideas toward automatic code generation.

%% file: introduction.tex


A rational software developer wants to make the most out of their time and work as efficiently as possible.
This is true in virtually every profession since no professional likes to waste their time or that of the
person or company that hires them. A developer that gets more work done in the same amount of time simply
generates more value, which is in the best interest of all parties involved.\\
This inherent interest in optimizing work efficiency that permeates all fields of work has been
one of the key motivations for many former and current research efforts that aim to 
assist developers in their day-to-day work and streamline the process of writing code.
~\cite{Bruch2009, Soeken2012, Chen2021, kim2021code, svyatkovskiy2020intellicode, pythia, Asaduzzaman2014}
\\
A major aspect of software development is dedicated to quality assurance (QA) and therein
most notably automated software testing.
Software testing plays a significant role in ensuring that the functional and non-functional
requirements of a system are met.
In other words, software testing makes sure a system behaves as expected.
Automation and code generation in the area of software testing reduce the manual
effort involved with QA which increases developer efficiency and enables systems
to scale better.~\cite{wang2020automatic, Escalona2011}
\\
There exist many methods and techniques in the realm of software testing, but in this
work, we will focus on behavior-driven development (BDD) tests. BDD is a practice that integrates natural language
(NL) into test specifications to make them comprehensible for people with little or no technical background.
\\
Initially developed for the processing of NL, recent progress in Artificial Intelligence (AI) and Machine
Learning (ML) enabled the application of novel techniques to coding assistance~\cite{Chen2021, kim2021code, svyatkovskiy2020intellicode, pythia}.
The transformer model has marked a milestone in natural language processing (NLP)\cite{AttentionIsAllYouNeed} and
is the technique we will be focusing on here but the ideas we present are not confined to any particular ML model.
\\
In this work, we propose an approach that further leverages the NLP capabilities of transformers for code generation
by using BDD test specifications as input for the generation task. We aim to provide developers with a tool to
optimize their efficiency while also enforcing established software engineering practices i.e. automated software
testing. While we explore and describe a possible implementation of this, we do not provide any proof-of-concept
solution as part of this work.

The rest of the paper is structured as follows: Section~\ref{sec:background} describes the research and technologies
our work is based upon and our reasoning behind choosing them. In Section~\ref{sec:idea} we will describe our approach
in detail. Section~\ref{sec:discussion} discusses the potential impact of our vision and Section~\ref{sec:conclusion}
gives a conclusion.
\raggedbottom

%% file: background.tex

\begin{figure}[h]
  \centering
  \includegraphics[width=1\columnwidth]{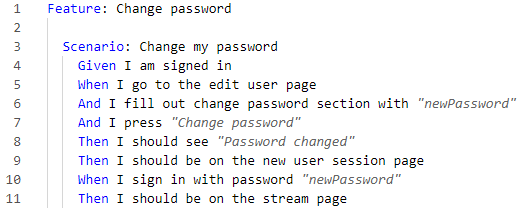}
  \caption{Cucumber Test specifying a change password feature, based on real-world example~\cite{diaspora}.}
  \label{fig:bdd-cucumber}
  \vspace{-2mm}
\end{figure}

As stated in~\cite{Trudova2020ArtificialII}, reducing the manual effort involved
with software testing is one of the key motivations driving current testing-related research.
The 2020 Literature Review by Trudova et al. has shown that most approaches focus on the
generation of test cases, while none attempt to generate code
from test specifications. This suggests that our described techniques are indeed novel.
We propose a method to generate application code from test case
specifications to further reduce the developer time needed in
application development. Specifically, we aim to extract the relevant information 
needed for the generation of code from the specification of test cases adhering
to behavior-driven development (BDD) standards.
\\
The practice of BDD is a technique that evolved from
test-driven development (TDD) \cite{bddVsTdd, bddBehaviourDriven, bddDanNorth}.
It is an agile software development process that, at its core, focuses on the
specification of test cases at the beginning of a development cycle together with
domain experts and stakeholders.
For the test cases to be interpretable
by these domain experts and stakeholders, the test case specification is usually
realized with a domain-specific language (DSL) that integrates natural language
and is executed by a specialized tool or framework (e.g. cucumber~\cite{cucumber}).
An example of a BDD test case for the cucumber framework can be found in
Figure~\ref{fig:bdd-cucumber}.

In 2012 Soeken et al.~\cite{Soeken2012} developed a technique to generate code
stubs from these BDD test case specifications using state-of-the-art NLP techniques
of the time. We propose a vision to take this idea further and generate functioning software
components from these natural language specifications using state-of-the-art
machine learning techniques like pre-trained transformer networks.
We have strong reason to believe that this approach is viable since transformers
have recently been shown to be highly effective in the generation of code from
natural language\cite{Chen2021} and the
processing of code and natural language in general~\cite{Feng2020, kim2021code, ESoTfSC}.
\\
In addition to the reduction of effort required in the development of applications,
we believe that our approach incentivizes the implementation of meaningful test
cases and therefore the adoption of QA practices which are
generally regarded as integral to the development of high-quality software.
\\
Having a general-purpose solution that can reliably generate application
code from BDD-test case specifications in an arbitrary context
(application type, programming language, libraries, etc.) would be ideal but
this defines a very complicated and nuanced research task which we do not
aim to address here and leave it up to future work. Instead, we want to focus on
the development of a system that works in a very defined context
to reduce overall system complexity and research whether the general approach is
feasible.
\\
We propose to build a prototype that works in the context of frontend development.
Virtually all applications developed in an enterprise scenario that are expected
to be used by multiple concurrent users (except for mobile applications) are web
applications. This means they are comprised of a backend running on one or more
servers and a frontend running inside of the client's browser. While it is possible
to build a frontend using plain HTML, JavaScript (JS), and CSS, this is very cumbersome
and most of the time developers opt for the use of a Framework when developing modern
frontends. Such frameworks ease the use of browser APIs and provide a way of structuring
applications in a reliable way while providing much utility to the developer.
For this research effort, we will be using the open-source Angular~\cite{angular} framework
developed by Google.

We base this decision on several reasons:
\begin{enumerate}
    \item \textbf{Popularity:}
    Angular has been on the rapidly changing market of JS frameworks for over two decades and
    has consistently been among the top 2 most used frontend frameworks in the past
    7 years.~\cite{frontend-popularity} This implies that there is presumably a large amount
    of publicly available, high-quality code in open-source repositories that can be used as
    training data for this project. Furthermore, this implies that reducing the development effort
    of angular applications would be relevant to many people and companies that rely on it
    today and most likely in the future.
    \item \textbf{Opinionated Framework:}
    In contrast to other frameworks like React~\cite{react},
    Angular is a very opinionated framework.
    This means that there exist many conventions
    regarding Angular applications resulting in the fact that the overall structure
    of most applications and components is roughly the same.
    Having less variance in the code artifacts that are to be generated should make it
    easier for the applied ML model to perform well, even with sparse training data.
    We assume this because this way the model does not need to pick up on multiple ways
    and approaches a problem can be solved but rather learn what is dictated by convention.
    %
    \item \textbf{Enforced BDD:}
    As Angular is very opinionated, it also imposes conventions
    around the testing methodology on users. When creating a new Angular component using
    the Angular CLI (which is generally the preferred way of doing so), a test file for
    the new component is automatically generated. All Angular tests are based on
    Jasmine~\cite{jasmine} which is a BDD framework for JavaScript and TypeScript.
    Because of this, we believe it is reasonable to assume that there exist many valuable
    samples of BDD specification and component code pairs in publicly available open-source
    projects that can be used for the training of our model. Also, because of this convention,
    we believe that the adoption of our proposed system should require minimal effort for
    a developer that is already accustomed to the Angular ecosystem.
\end{enumerate}

To process the acquired data and use it for ML, we need an ML model that trains on the
provided data so that it eventually becomes capable of generating the desired output given an
input.
We propose an approach that uses transformers~\cite{AttentionIsAllYouNeed}.
Transformers are a relatively novel alternative to other deep neural network architectures such as 
convolutional (CNN) and recurrent neural networks (RNN) that promises benefits in terms of 
scalability and performance. 
Recent work has shown that transformers can be adopted for code generation and prediction with
the results outperforming the state-of-the-art.~\cite{Feng2020, Chen2021, kim2021code, svyatkovskiy2020intellicode}
\\

%% file: method.tex

As our proposed system will be built using a state-of-the-art ML model, we require sample data
for training, testing, and validation. Details about how we plan to acquire and pre-process
this data can be found in Section~\ref{ssec:data_aquisition}. Details about the ML model can be found in Section~\ref{ssec:model}. The intricacies of our plans
for the code generation process can be found in Section~\ref{ssec:post_processing}.
With our proposed approach, we plan to answer the following research questions:
\begin{itemize}
\item
    \textbf{RQ1}:
    What is the best way to curate data to train a model on angular test cases?
    %
    \item \textbf{RQ2}:
    Can our proposed approach generate high-quality code?
    %
    %
    \item \textbf{RQ3}: 
    How high is the instruction/branch coverage of generated application code?
    %
    %
\end{itemize}

The proposed system will be evaluated to find answers to the aforementioned
research questions which is further elaborated in Section~\ref{ssec:evaluation}.

\subsection{Data Aquisition and Pre-Processing}
\label{ssec:data_aquisition}

\begin{figure}[h]
  \centering
  \includegraphics[width=1\columnwidth]{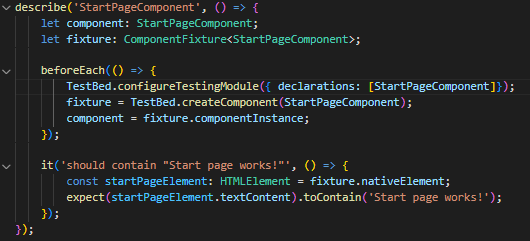}
  \caption{Example Angular component test}
  \label{fig:bdd}
  \vspace{-2mm}
\end{figure}

As stated at the beginning of this section, we need data
for the training, testing,  and validation of our model and we plan to extract
this data from open-source projects on GitHub.
\\
There are several considerations to be made during the acquisition of data in this
specific context. First of all, not all parts of an angular application are created equal.
In general, code containing files in an Angular project can be of different types. This 
includes the non-exhaustive list of \textit{Components}, \textit{Services}, and 
\textit{Modules}.
Explaining the function of each of these file types is beyond the scope of this
paper. Here we provide a brief explanation.
\textit{Components} are where usually most of the application code resides.
A \textit{Component} describes
a piece of user interface along with its markup, styling, and logic that can be arranged
hierarchically. \textit{Services} are singleton objects that make data and
functionality
available inside of a \textit{Module}. \textit{Modules} (as the name suggests) define code modules
that encapsulate multiple \textit{Components}, \textit{Services}, etc. in a defined scope.
\\
Being able to generate all of these different parts of an application is very desirable
but for our approach, we propose to first keep the complexity as low as possible
and only focus on the generation of \textit{Components} and establish a baseline to incrementally build upon. If the first
results appear promising then one could think about extending the system to also be
able to generate \textit{Services} and eventually all other application parts.
We focus on the generation of \textit{Components} since they represent the basic building block
in angular applications. It is possible to build an angular application solely out of
\textit{Components} which is not true for any of the other application parts.
\\
Filtering the scraped code to only contain \textit{Component} definitions is
trivial to implement because by convention all \textit{Component} files are named
\texttt{<COMPONENT\_NAME>.component.ts} and their accompanying
test specifications
\texttt{<COMPONENT\_NAME>.component.spec.ts}.
Using a regular expression, picking up on this naming convention is straightforward
and yields a set of \textit{Component} definition and test specification pairs.
An example of a component test with one test case can be found in Figure~\ref{fig:bdd}.

A first analysis has shown that using the described method $\sim 2.5M$
angular \textit{Component} code and test pairs can be found on GitHub which is a
promising first result. Details of this analysis can be found in
Table~\ref{tab:github_data}.

\begin{table}[h]
    \centering
    \begin{tabular}{lr}
    \hline
     Type                        &        \# in thousands \\
    \hline
     Angular Repositories    &   383 \\ 
     Component Test Pairs &  2 547 \\ 
     Service Test Pairs   &  1 312 \\ 
    \hline
    \end{tabular}
    \vspace{2mm}
    \caption{First analysis of potential training data on Github}
    \label{tab:github_data}
      \vspace{-6mm}
\end{table}

To ensure that the data is of high quality, we propose to filter the data further and 
remove samples from the data set that do not meet certain criteria. Possible exclusion
criteria could be the number of test cases per test file (e.g. minimum of 3 test cases)
or the amount of forks of the associated GitHub project (e.g. minimum of 5 forks).
We believe these filters to be promising, since angular auto generates one test case per
\textit{Component} and we are of the opinion that more are needed for the \textit{Component} to
be considered high quality.
Also we assume that GitHub users are more likely to interact with a project if they think
it contains high quality code.
Additionally, it is most likely beneficial to bring all of the data into a standardized form.
Angular component definitions are oftentimes split up into different files each
containing only either markup, styling or logic. By inlining the markup and styling into
the logic part where necessary, we can omit this issue.
\\
Several research efforts have shown promising results in ML code processing tasks when
the code is converted to tokens or an AST~\cite{allamanis17}.
Depending on the model that is ultimately used, this
is something that we propose to explore.
Since Angular 2+ code is always written
in TypeScript~\cite{typescript} which is a compiled language that has open source
tooling, the tokenization and AST conversion of our data can be achieved with
production-ready, publicly available solutions.

%
%


%
%
%
%

\subsection{Model}
\label{ssec:model}
At the core of our proposed approach is a generative ML model and because
of the recent ubiquity and performance of transformer-based models in generation
and sequence-to-sequence transformation tasks~\cite{AttentionIsAllYouNeed, Chen2021, Feng2020, kim2021code, ESoTfSC},
using a transformer in our described scenario becomes an obvious choice.
\\
Because of the comparatively low effort involved, as a first approach, we propose
to use OpenAI's Codex model which is based on GPT-3\cite{Chen2021}. Codex has been used in
very successful code generation applications like GitHub's Copilot~\cite{copilot}. Because
it is pre-trained on a vast amount of natural language and code and implements
a high-performing architecture, using Codex can potentially reduce the development
time of our system by a very large amount when compared to
implementing and training a custom model.
Through the use of the Codex fine-tuning API, our data set can be used to train
the model and customize it to our needs.
\\
Additionally, we suggest the use of CodeBERT~\cite{Feng2020} and CodeT5~\cite{codeT5} which
are other state-of-the-art pre-trained transformer models designed and
trained for the processing and generation of code. Comparing the performance of these
models will likely give further insights into which style of transformer-based model
is best suited for this type of generation task.

\subsection{Post-Processing}
\label{ssec:post_processing}

Because we will use test case specifications as the input to our generation task,
we have a high amount of information regarding the desired output. We can leverage
this and have our model generate code samples which we then execute the tests
on in a sandboxed environment. Generated samples that do not meet all assertion criteria of our tests will be
discarded and only the ones that do are presented to the user. Since the
test implementations of our input data can be used in this way and jasmine tests are
comprised of strings containing NL and the test code as Figure~\ref{fig:bdd} shows, we propose to
explore the idea of only using the NL part of the test specifications
as the input to our model and use the test implementation for the selection of
generated results.
\\
When generating Angular \textit{Component} code, there will most likely be issues with
\textit{Component} dependencies. Angular relies on an Inversion of Control (IoC) mechanism, specifically Dependency Injection. It is a common pattern to
have functionality or data contained in a \textit{Service} which is then injected
into multiple \textit{Components} where the functionality or data is then accessed.
When \textit{Component} code is generated for a \textit{Component} that depends on
another part of the application that is not implemented yet, compiler errors
will arise when the execution of the tests is prepared.
This might be the case even though the generated code is correct and this dependency
is desired by the user. To combat this issue, we propose that in another post-processing
step code stubs for unmet dependencies are created. 
This means that if e.g. a dependency on a \textit{Service} that does not
exist yet is generated
by our model, then a stub for this \textit{Service} will also be created.
Similarly, if a dependency
on a method of a \textit{Service} is generated and the \textit{Service} exists but the method
does not, a stub for the method is created in the existing \textit{Service}.\\
The realization of this functionality should be straightforward because Angular
provides powerful tooling for the programmatic creation and modification of parts
of an existing application with Angular Schematics~\cite{schematics}.

\subsection{Evaluation}
\label{ssec:evaluation}

To answer our posed research questions, the proposed system needs to be evaluated
therefore we'll describe possible evaluation strategies in the following section.
\\
\textbf{RQ1} and \textbf{RQ2} are concerned with the quality of the generated
code but assessing the quality of such generation results is no easy task.
Ren et al.~\cite{Ren2020} show that match-based metrics like the BLEU score are
insufficient for the evaluation of generative
models for code since they have problems capturing semantic features specific to code.
To evaluate the output quality of our proposed system the CodeBLEU score
described by Ren et al. could be used but it had to be extended to be usable with
TypeScript. Additionally, an evaluation set of input-output pairs would need to be
created, similar to the HumanEval set in~\cite{Chen2021}.
Another possible evaluation strategy would be to have a representative group
of experienced developers that assess the quality of the generated code manually.
\\
To find a satisfactory answer to \textbf{RQ1} we propose a search over the
pre-preprocessing and filter parameters (Section~\ref{ssec:data_aquisition}) training
multiple models and evaluating code quality in one of the ways mentioned above.
\\
To find an answer to \textbf{RQ2} we suggest defining a threshold score that indicates
whether a piece of code can be considered high quality when measured against a sample from
the evaluation set. Whether the proposed system can reliably generate such code answers
the research question.
\\
Answering \textbf{RQ3} is straightforward since test coverage analysis is part
of the angular tooling. Such analysis could be run after the generation and the results
aggregated to calculate min, max, and average branch and instruction coverage on
test data.

%% file: impact_analysis.tex
In previous chapters, we have explained in detail how our proposed idea can be executed.
We will now highlight the potential value of this work.
\\
If enough high-quality data could be mined (\textbf{RQ1}) to successfully train an ML model that can generate high-quality code (\textbf{RQ2}) the viability of our approach would be demonstrated. Because of Angular's popularity,~\cite{frontend-popularity} this would imply potentially huge time and therefore cost savings for many people and companies.
\\
Additionally, it would open the door to further research that tries to achieve something similar
on another platform.
An interesting aspect that emerges from the fact that we rely on BDD tests is
that in an enterprise scenario, the BDD test cases often define the acceptance
criteria for user stories of the stakeholders. This implies that if our approach
can successfully generate application code that satisfies the tests, the code
has already passed quality assurance and is accepted from a business perspective,
theoretically eliminating all initial programming work that deals with the implementation
of actual application code.
\\
But even if the BDD test cases do not directly correspond with acceptance criteria, we theorize that
there are more benefits to our approach than simple time-saving. During development, software developers
usually take the most time-efficient route when solving a problem. This is especially true when
the project is on a tight schedule, as they oftentimes are. What appears as the most time-efficient route
sometimes does not correspond with the optimal route from a software engineering standpoint.
The result of this is that sometimes developers do not focus on writing enough meaningful tests, since
tests do not directly contribute to application functionality.
If our approach succeeds in generating high-quality code from test specifications, developers would be
provided with a very time-efficient way of solving their problems fast that requires them to write tests.
Additionally, we theorize that higher quality output can be generated by our proposed method when higher
quality input i.e. test specifications are provided which would further incentivize developers to write
more meaningful tests.
In case our approach yields equal or higher test coverage (\textbf{RQ3}) than what is generally achieved
in Angular projects, one could argue that the adoption of our techniques tends to increase
software quality which is a very desirable trait.\\
Furthermore, we theorize that using just the natural language section of the BDD tests will generate better results than using the tests in their entirety. Even if our theory is ultimately wrong, knowing the answer to this question will be beneficial for other researchers in the field potentially saving work or sparking research in different directions.\\

%% file: conclusion.tex
The field of ML-based code generation is just emerging and to our knowledge, there does not exist any research
that tries to achieve what we propose to do.
We describe a possible method for generating Angular frontend components based on BDD test specifications
by utilizing ML techniques.
Our approach aims to reduce the development time needed for web applications and introduce new research
ideas toward automatic code generation. \\
From our impact analysis, we believe that this approach has the potential to bring many benefits to academia and industry, which is why we advocate for further research in this direction.

%% file: acknowledgement.tex
This research work has been funded by the German Federal Ministry of Education and Research
and the Hessen State Ministry for Higher Education, Research and the Arts within their joint
support of the National Research Center for Applied Cybersecurity ATHENE and Crossing SFB.